\documentclass[prl,showpacs,twocolumn]{revtex4}
\usepackage{epsfig}

\begin{document}


\title{Dimensional Control  of Antilocalization and Spin Relaxation in Quantum Wires}

\author{ S.  Kettemann}

\affiliation{ I. Institut f. Theoretische Physik, Universit\" at Hamburg,
 Hamburg 20355,  Germany}



\begin{abstract}
  The spin relaxation rate $1/\tau_s (W)$ 
  in disordered   quantum wires with 
    Rashba and Dresselhaus spin-orbit coupling is derived 
    analytically as a function of  wire width $W$.
     It is   found to be   diminished 
    when  $W$ is smaller than 
     the bulk spin-orbit length  $L_{\rm SO}$. 
     Only  a small  spin relaxation rate 
       due to  cubic Dresselhaus coupling $\gamma$ is found to remain in this limit.
  As a result, when reducing the wire width $W$   the quantum  conductivity correction  
        changes   from weak anti- to weak localization and  from negative to  positive magnetoconductivity.
\end{abstract}
\maketitle 
 Quantum interference of electrons in low-dimensional, disordered conductors
    results in corrections to the  electrical 
  conductivity $\Delta \sigma$.
     This quantum correction, the    weak localization 
   effect, is known to be 
    a very sensitive tool to study 
     dephasing and symmetry breaking mechanisms  in conductors\cite{review}. 
  The
      entanglement of spin and charge by   spin-orbit interaction
      reverses the effect of weak localization and 
      thereby enhances the conductivity, the weak antilocalization effect.
   Since the electron  momentum is   randomized due to disorder,
    spin-orbit interaction  results
     in   randomization of the 
    electron spin,
         the  Dyakonov-Perel  spin relaxation with rate $ 1/\tau_s$ \cite{perel}. 
         This spin relaxation 
         is expected to vanish
 in   narrow wires
  whose width $W$  is of the order of  Fermi wave length $\lambda_F$\cite{kiselev,meyer}. 
  In this article we  show, however,  that 
  $1/\tau_s$ is already 
  strongly reduced in wider  wires:
  as soon as the wire  width $W$ is smaller than  
  bulk spin-orbit length  $L_{\rm SO}$. 
  This  explains the 
   reduction of spin relaxation rate in 
n-doped    InGaAs-wires, as recently observed with  optical  \cite{holleitner} 
   as well as with  weak localization measurements \cite{hu05,gh05,schaepers,lshh04}. There, 
   $L_{\rm SO}$  is  
   as large as  several  $\mu m$, and exceeds both the elastic mean free path $l_e$, and  $\lambda_F$. 
  In clean, ballistic 2D electron systems (2DES), $L_{\rm SO}$ is 
  the length on which the  electron spin     precesses
    a full cycle.
 It is important to note that this length scale is not changed  as  the   wire width  $W$ is reduced below $L_{SO}$, because
  the spin orbit interaction remains of the same order as in 2D systems.  
 Therefore, this reduction of spin relaxation has the following important consequence: 
    the spin of  conduction electrons 
can   precess coherently as it moves along the wire on  length scale $L_{SO}$.  
 The spin   becomes
   randomized  and  relaxes  on the  longer 
    length scale  $L_s (W) = \sqrt{D \tau_s}$, only
    ($D= v_F^2 \tau/2$ ($v_F$, Fermi velocity) is the 2D diffusion constant). 
    Therefore,  the dimensional reduction 
     of  spin relaxation rate  $1/\tau_s (W)$ can be  
     very useful for the realization of spintronic devices,
   which rely on  coherent spin      		evolution\cite{dasdatta,spintronics}.

 Weak antilocalization
  was  predicted  by Hikami, Larkin, and Nagaoka \cite{nagaoka}
  for conductors with  impurities of heavy elements.  As   conduction 
    electrons scatter from such impurities, 
     the   spin-orbit interaction 
     randomizes their spin.
 The resulting   spin relaxation suppresses   interference  of  time reversed paths  in  spin 
  triplet configurations, 
  while  interference in  singlet  configuration
    remains unaffected. Since  singlet interference
       reduces the electron's
      return probability it  enhances the conductivity, 
       the weak antilocalization effect. 
  Weak magnetic fields  suppress  the singlet 
   contributions, reducing    the conductivity  and 	resulting in negative magnetoconductivity.
  If the   host lattice of the electrons provides 
  spin-orbit interaction,  quantum corrections
 to the conductivity have to be calculated in the basis of 
 eigenstates 
 of the Hamiltonian  with 
 spin-orbit interaction,   
\begin{equation} \label{hamiltonian} 
H_0 = (\hbar^2/2 m_e) {\bf k^2} + \hbar  \mbox{\boldmath$\sigma$} {\bf  \Omega}, 
\end{equation}
  ($m_e$, effective electron mass),  ${\bf \Omega}^T = (\Omega_x, \Omega_y) $, are
    precession frequencies of the electron spin 
 around the x- and  y-axis. {\boldmath$\sigma$} is a 
  vector, with components 
  $\sigma_i$, $i =x,y$,  the Pauli matrices.
 The breaking of  inversion symmetry 
 causes  a spin-orbit interaction, 
 given by 
 \cite{dresselhaus}
\begin{equation}
{\bf \Omega_D } = \alpha_1 ( -\hat{e}_x k_x + \hat{e}_y k_y)/\hbar
+ \gamma  ( \hat{e}_x k_x k_y^2 - \hat{e}_y k_y k_x^2)/\hbar.
\end{equation}
  $\alpha_1 = \gamma \langle k_z^2 \rangle$, the linear Dresselhaus parameter,
   measures the strength of the term  linear in    momenta $k_x, k_y$ in the plane of the 2DES.
  When  $\langle k_z^2 \rangle \sim 1/a^2 \ge k_F^2$  ($a$,  thickness of the 2DES, 
    $k_F$, Fermi wave number), that term  exceeds the cubic 
    Dresselhaus terms with coupling $\gamma$. 
 Asymmetric confinement 
 of the 2DES yields  the  Rashba term ($\alpha_2$,Rashba parameter) \cite{rashba},
\begin{equation}
{\bf \Omega_R} = \alpha_2  ( \hat{e}_x k_y - \hat{e}_y k_x)/\hbar.
\end{equation}
  The quantum correction to the conductivity $\Delta \sigma$ arises from the fact, that the 
   quantum return probability to a given point ${\bf x_0}$ 
      after a time $t$, $P(t)$,   differs from 
    the classical return probability,
due to      quantum interference. 
   Therefore, $\Delta \sigma$
     is proportional to 
     a time  integral over  the quantum mechanical return probability $P (t) = \lambda_F^d(t) n({\bf x_0},t)$
       ($d$,   dimension of diffusion, 
         $n$,  electron density). 
    For uncorrelated disorder potential, $V(x)$, with  
      $\langle V \rangle=0$ and $\langle V(x) V(x')\rangle = \delta (x-x')/2 \pi \nu \tau$ 
      ($\nu = m/(2 \pi \hbar^2)$,  average density of states per spin channel, 
        $\tau$,  elastic mean free time), we can   
    perform the disorder average. Going 
     to momentum (${\bf Q}$) and frequency
      ($\omega$)  representation, and 
      summing up  ladder diagrams to take into account  the  diffusive motion, yields
    the quantum correction to the static conductivity \cite{nagaoka},
    \begin{equation}
    \Delta  \sigma
    = -  \frac{2 e^2}{h}\frac{\hbar D}{ Vol.} \sum_{\bf Q} \sum_{\alpha, \beta = \pm} C_{\alpha \beta \beta \alpha,  \omega=0} ({\bf Q}),
    \end{equation}
   where $\alpha,\beta = \pm$ are the spin indices,  and   the Cooperon propagator $\hat{C}$  is
       for $ \epsilon_F \tau \gg 1$ ( $\epsilon_F$, Fermi energy),
       and  neglecting the Zeeman coupling, 
       \begin{eqnarray}
      \hat{C}({\bf Q})^{-1}&=&  \frac{\hbar}{\tau}  - \int \frac{d \Omega}{\Omega}
       \frac{\hbar/\tau}{ 1  + i \frac{\tau}{\hbar} {\bf v} ( \hbar {\bf Q} +2 e {\bf A} + 2m_e \hat{a} {\bf S} ) }.
       \end{eqnarray}
       The  integral is over all angles of  velocity   $ {\bf v}$ on the 
        Fermi surface ($\Omega$,  total angle.
       $e$, electron charge, $ {\bf A} $, vector potential). 
        ${\bf S}$ is the total spin vector
         of  spins of time reversed paths:
       $
       {\bf S} =  \left(\mbox{\boldmath$\sigma$} + \mbox{\boldmath$\sigma$}' \right)/2 
       $. 
         $\hat{a}$ is the 2 by 2 matrix
     \begin{equation}
\hat{a} = \frac{1}{\hbar}  \left(   \begin{array}{cc}
-\alpha_1 + \gamma k_y^2& - \alpha_2 \\
\alpha_2 & \alpha_1 - \gamma k_x^2
\end{array} \right).
     \end{equation}
       In 2D, the angular integral  is continuous from $0$ to $2 \pi$, yielding  to lowest order in  $( {\bf Q} +2 e {\bf A} + 2m \hat{a} {\bf S} )$, 
      \begin{equation} \label{cooperon2}
      \hat{C} ({\bf Q})= \frac{\hbar}{D (\hbar {\bf Q} + 2 e {\bf A} + 2 e  {\bf A}_{\bf S})^2 + H_{\gamma}  }.
  \end{equation} 
      The effective vector potential due to spin-orbit interaction,
  ${\bf A}_{\bf S} = m_e  \hat{\alpha} {\bf S}/2$,   ($
       \hat{\alpha} = \langle\hat{a}\rangle$)  couples 
       to  total spin $ {\bf S}$.
     The cubic Dresselhaus coupling reduces the effect of the linear one 
       to
      $\alpha_1 - m_e \gamma \epsilon_F/2$. Furthermore, 
      it gives rise to  the
         spin relaxation term in Eq. (\ref{cooperon2}), 
      \begin{equation} \label{hgamma}
      H_{\gamma} =  D \frac{m_e^2\epsilon_F^2 \gamma^2}{\hbar^2} (S_x^2 + S_y^2).
      \end{equation}
    In the representation of the singlet,  
       $\mid~S=0;m=0\rangle$
       and triplet states $\mid~S=1;m=0,\pm\rangle$,
      $\hat{C}$  decouples into a  
       singlet and a triplet sector. Thus, the  quantum conductivity is  a 
        sum of  singlet and triplet terms, 
       \begin{eqnarray} \label{qmc}
    \Delta  \sigma
   & = &- 2 \frac{e^2}{h}\frac{\hbar D}{ Vol.} \sum_{\bf Q} \left( 
    - \frac{\hbar}{D (\hbar {\bf Q} + 2 e {\bf A})^2} \right. \nonumber \\
     && \left. 
     +
     \sum_{m=0,\pm1}  \langle S=1,m \mid \hat{C} ({\bf Q}) \mid S=1,m\rangle  \right).~~~~~
    \end{eqnarray}
    The triplet terms have been  evaluated in various approximations  before\cite{knap,miller,af01,lg98,golub}.          
In 2D   one can  treat the magnetic field nonperturbatively, 
 using  the basis of  Landau bands\cite{nagaoka}. 
 In wires with   widths smaller than  cyclotron length $k_F l_B^2$ 
  ($l_B$, the magnetic length,  defined by $ B l_B^2 = \hbar/e$),  the Landau basis is not suitable. 
Fortunately,  there is another way  to treat  magnetic fields: 
   quantum corrections are due to the interference between closed time reversed paths.  In     magnetic fields the electrons  acquire a magnetic phase, which breaks  time reversal 
   invariance.  Averaging over all  closed  paths, 
  one obtains a   rate with which the 
 magnetic field breaks the time reversal invariance,
  $1/\tau_B$.
 Like  the dephasing rate $1/\tau_{\varphi}$, it cuts off 
 the divergence arising from  quantum corrections with  small
  wave vectors ${\bf Q}^2 < 1/ D \tau_B$.
 In 2D  systems, $\tau_B$ is  the time an electron  diffuses 
  along a closed path   
 enclosing  one magnetic flux quantum,      $ D \tau_B = l_B^2$.
 In wires of finite width $W$
  the area which the  electron path  encloses  in a time 
 $\tau_B$ is   $ W \sqrt{D \tau_B} $. Requiring that 
 this   encloses one flux quantum gives
$1/\tau_B = D e^2 W^2 B^2/(3 \hbar^2)$.
For arbitrary  magnetic field  the relation
$ 1/\tau_B= D (2 e)^2 B^2 \langle y^2 \rangle/\hbar^2$  with the  expectation value of the square of the transverse position $\langle y^2\rangle$, yields 
$1/\tau_B=D/l_B^2\left(1-1/(1+W^2/3l_B^2)\right)$.
 Thus,  it is sufficient to  diagonalize  the Cooperon 
 propagator as given by  Eq.(\ref{cooperon2}) without magnetic field and 
  to add   the magnetic 
 rate $1/\tau_B$  together with  dephasing rate $1/\tau_{\varphi}$ to the denumerator of $\hat{C} ({\bf Q})$, when calculating the conductivity correction,  Eq. (\ref{qmc}).  
  
 It is well known that  the Cooperon propagator can be diagonalized
 in 2D for pure Rashba coupling $\alpha_1=0,\gamma=0$, or pure Dresselhaus coupling 
    $\alpha_2=0$\cite{iordanskii,knap,lg98,golub}. For example,  keeping only  Rashba coupling $\alpha_2$ 
  the three triplet Cooperon Eigenvalues   are in 2D,   
   \begin{eqnarray} \label{triplet2D}
    &&  E_{T 0}/(D \hbar) =  {\bf Q}^2 +  Q_{SO}^2, 
   \nonumber \\  
   && E_{T \pm}/(D \hbar) =  {\bf Q}^2 +  \frac{3}{2} Q_{SO}^2 
   \pm \frac{1}{2} Q_{SO}^2 \sqrt{1 + 16 \frac {{\bf Q}^2}{Q_{SO}^2 } }, ~~~~~
   \end{eqnarray}
   where $Q_{SO} = 2 m_e \alpha_2/ \hbar^2$.
   If we use the  approximation,  
   \begin{equation}\label{triplet2D2}
   E_{T \pm}/(D \hbar) \approx  ( Q \pm Q_{SO})^2 +   Q_{SO}^2/2,
   \end{equation}
    which 
    is  plotted for comparison with the exact dispersion, Eq. (\ref{triplet2D}) in Fig. 1,
    we can   
 integrate analytically  over the 2D momenta. Thus,    the 2D quantum   correction is
 \begin{equation}  \label{wl2D}
 \Delta \sigma = - \frac{1}{2 \pi}\ln \frac{H_{\varphi}}{H_{\varphi} +  H_{s}}  
 +   \frac{1}{\pi} \ln  \frac{H_{\varphi} +  H_{s}/2}{ H_{\tau} },
 \end{equation}
 in units of $e^2/h$.
 All parameters are rescaled to dimensions of magnetic fields: 
 $H_{\varphi} = \hbar/(4 e D \tau_{\varphi})$,  $H_{\tau} = \hbar/(4 e D \tau)$,
  and  the spin relaxation field
  $ H_{s} = \hbar/(4 e D \tau_{S xx})$\cite{knap}.
   The 2D  spin relaxation rate of one spin component  is
    for pure Rashba coupling, 
    $ 1/\tau_{S xx} = 1/\tau_s = 2 p_F^2  \alpha_2^2 \tau$  \cite{knap,iordanskii}, and 
      is related to spin-orbit gap 
       $\Delta_{SO}=   \hbar v_F Q_{SO}$, by   $1/\tau_s =  (\Delta_{SO}/\hbar )^2     \tau/d $.     

       Note that the magnetoconductivity is dominated by the minima of the dispersion
        of Cooperon eigenvalues. Therefore, these minima,
         whose finite value  we may  call spin relaxation gaps, 
      are  a direct measure of  spin relaxation rate. 
          We note that the lowest minima  
    of the  triplet modes are shifted to nonzero 
       wave vectors, $Q = \pm Q_{SO}$. Thus, the spin relaxation gap 
        is   by  about 
        a   factor $1/2$ smaller, than at $Q=0$
         \cite{iordanskii}.
     
    Without spin-orbit interaction, 
     the   conductivity of quantum wires 
      with width $W < L_{\varphi}$ is dominated by  the transverse zero mode
        $Q_y =0$.  
  This   
     yields  the quasi-1D weak localization correction  as
  used previously for narrow GaAs wires\cite{kurdak}. 
 However,  in the presence of spin-orbit interaction,
setting simply $Q_y =0$ is not correct.
 Rather one  has  to 
  solve the Cooperon equation
  with the modified  boundary conditions\cite{af01,meyer},    
\begin{equation} \label{bc}
(-i \partial_y + 2 e  A_{S y} ) C( x, y = \pm W/2) = 0, 
\end{equation}
for all $x$. 
     \begin{figure} [t]
\begin{center}
\vspace{-.5cm}
\epsfig{file=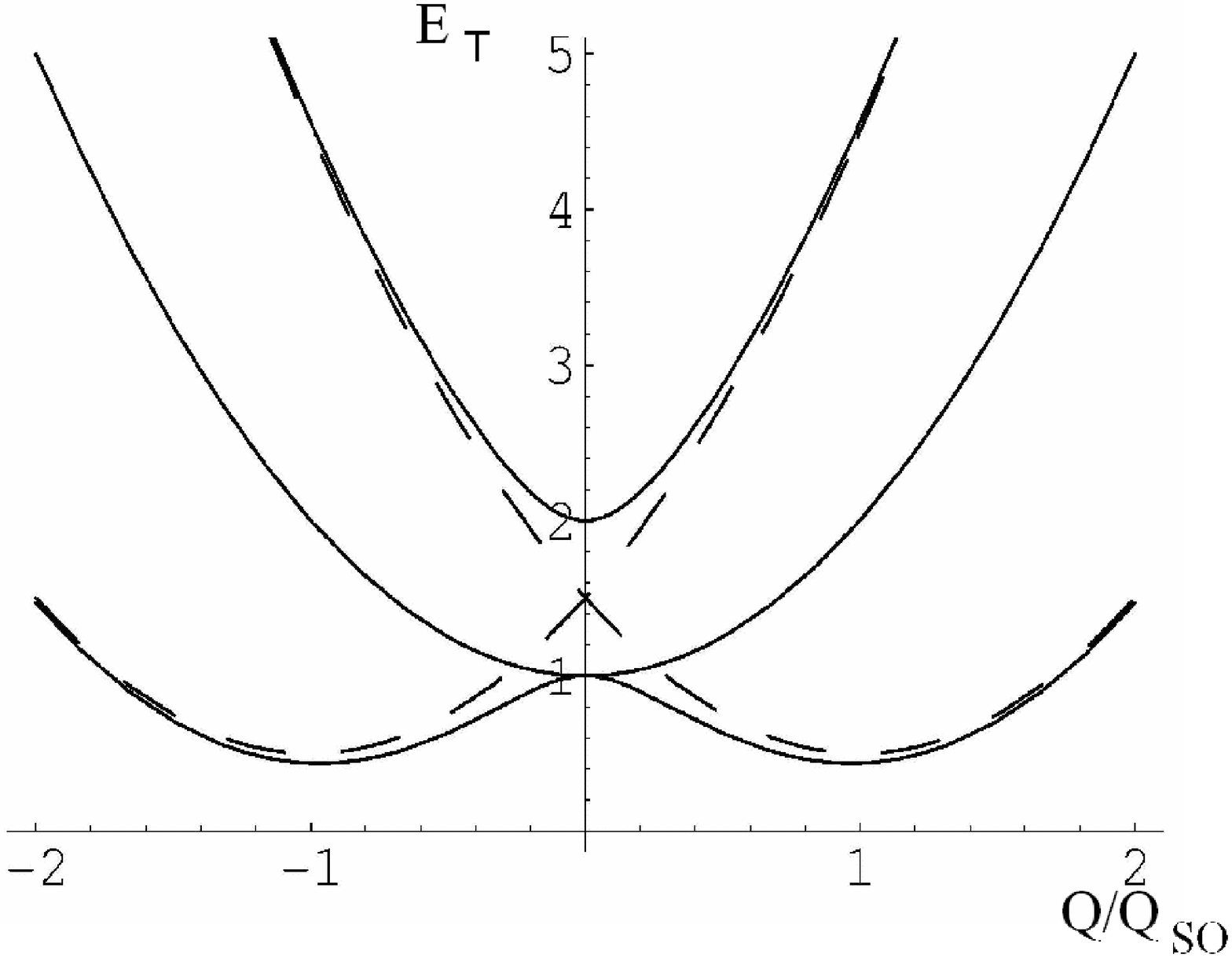,width=5cm} \caption{ 
 Dispersion of   triplet Cooperon modes in 2D  in units of $\hbar D Q_{SO}^2$, Eqs. (\ref{triplet2D}) (full lines), and 
  Eq. (\ref{triplet2D2}) (dashed lines).
 } \label{Fig1}
\vspace{-.9cm}
\end{center}
\end{figure}
 The transverse zero mode $Q_y =0$ does not satisfy this condition.
Therefore,   it is convenient to  perform a Non-Abelian 
  gauge transformation   \cite{af01}. 
  Since in quantum wires   these boundary conditions apply 
  only in the transverse       direction,
    a transformation  in the transverse  direction is needed, only:   $\hat{C} 
  \rightarrow \tilde{C} = U \hat{C}  \bar{U}$, with
    $U = \exp (i 2 e A_{S y} y/\hbar)$. Then,   the
     boundary condition simplifies to,  $-i \partial_y \tilde{C}(x,y=\pm W/2)~=~0$. 
    For $ W < L_{\varphi}$  we can  use  the fact that  transverse nonzero modes
     contribute terms to the conductivity which are 
       a factor $W/n L_{\varphi} $ smaller than  the 0-mode term, 
        with $n$ a nonzero integer number.  
     Therefore,  it is sufficient to diagonalize the 
       effective quasi-1-dimensional
 Cooperon propagator: the transverse 0-mode expectation value 
  of the   transformed inverse Cooperon propagator $\tilde{H}_{1D} = \langle0\mid \tilde{C}^{-1}\mid 0\rangle$.
    It is crucial to note that  additional terms are created in 
 $\tilde{H}_{1D}$  by  the non-Abelian  transformation. 
 We can   diagonalize $\tilde{H}_{1D}$,  neglecting  small relaxation due to cubic Dresselhaus coupling  $\gamma$. 
We introduce the notation,  $Q_{SO}^2 = Q_D^2 + Q_R^2$ where 
  $Q_D$  depends
  on     Dresselhaus   spin-orbit coupling, $Q_D =  m_e (2 \alpha_1  - m_e \epsilon_F \gamma )/\hbar $. 
     $Q_R$ depends on   Rashba coupling: 
        $Q_R = 2 m_e \alpha_2/\hbar$.
  We finally   find  the 
 dispersion  of  quasi-1D triplet modes,  
  \begin{eqnarray}\label{e0}
 && \frac{E_{T 0}}{\hbar D} =  Q_x^2 +  Q_{SO}^2 \delta_{SO}^2 \left( \frac{1}{2} t_{SO} \delta_{SO}^2 
   + 2 c_{SO} 
   (1- \delta_{SO}^2)  \right),
   \nonumber \\ 
  &&  \frac{E_{T \pm}}{\hbar D} =   Q_x^2 +\frac{1}{4}  Q_{SO}^2 \left( 4 -  t_{SO} \delta_{SO}^4   - 4 c_{SO} \delta_{SO}^2  (1-\delta_{SO}^2)
   \right.  \nonumber \\  && \left.
      \pm 2 \sqrt{ h(\delta_{SO}) +   \frac{16 Q_x^2}{Q_{SO}^2} (1+c_{SO}(c_{SO}-2) \delta_{SO}^2)}\right),
  \end{eqnarray}
  where 
$\delta_{SO} = (Q_R^2-Q_D^2)/Q_{SO}^2$, and 
  \begin{equation}
  c_{SO} = 1-  \frac{2 \sin (Q_{SO} W/2)}{Q_{SO} W},~
 t_{SO} = 1-  \frac{ \sin (Q_{SO} W)}{Q_{SO} W}.
  \end{equation}
 Here,  $h(\delta_{SO}) =t_{SO} \delta_{SO}^8/4   +
    \delta_{SO}^2 (1-\delta_{SO}^2)(
  4 c_{SO}^2 (1- 3 \delta_{SO}^2 + 3 \delta_{SO}^4) + t_{SO}^2 \delta_{SO}^2 (1+\delta_{SO}^2) - 6 c_{SO} t_{SO} \delta_{SO}^4)$.
   In Fig. \ref{Fig2}, the gap of $E_{T 0}$ and the full dispersion 
    of the other two triplet modes are plotted for pure Rashba coupling $\delta_{SO} =1$, 
     as a function of the wire width $W$ as scaled
      with $Q_{SO}$.
       In Fig. \ref{Fig5} the magnetoconductivity is plotted  for pure Rashba coupling
    $\delta_{SO} =1$     as
        function of the wire width $W$. Inserting Eq. (\ref{e0}) 
         into the expression for the quantum correction to the conductivity Eq. (\ref{qmc}),  
          the integral over momentum $Q_x$ is done  numerically. We note a change of sign from weak antilocalization 
          to weak localization as $Q_{SO} W$ becomes smaller than 1. In the crossover regime
       $Q_{SO} W \approx 1$    very weak magnetoconductivity
           is  found. 
\begin{figure}
 \begin{center}
 \vspace{-.5cm}
\epsfig{file=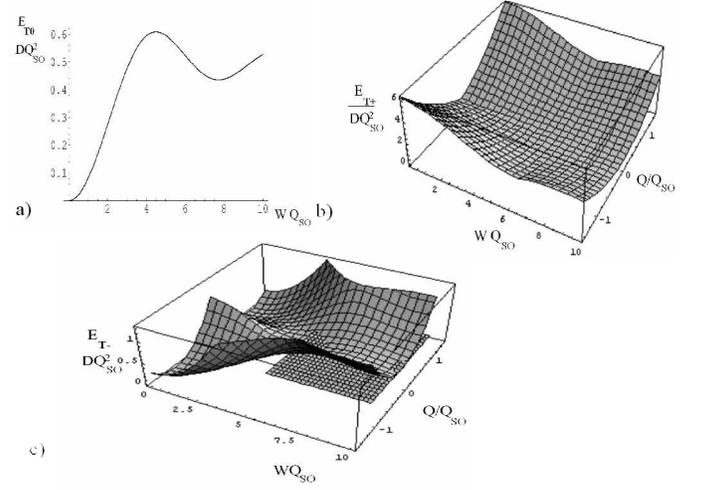,width=9cm} 
\vspace{-.5cm}
\caption{
 For  pure Rashba coupling  $\delta_{SO}=1$: a) Gap of   Triplet mode $E_{T 0}$  as function 
  of wire width   $W$ (in units of $L_{SO} = 1/Q_{SO}$). 
  b)  Dispersion of    Triplet mode $E_{T +}$ and 
   c)  of   $E_{T -}$  as function 
  of  width  $W$ and momentum 
   $Q$ (scaled with  $Q_{SO}$) and  $E/(\hbar D Q_{SO}^2)= 1/2$
    for comparison. 
  } \label{Fig2}
  \vspace{-.5cm}
\end{center}
\end{figure}
\begin{figure}
\begin{center}
\vspace{-.5cm}
\epsfig{file=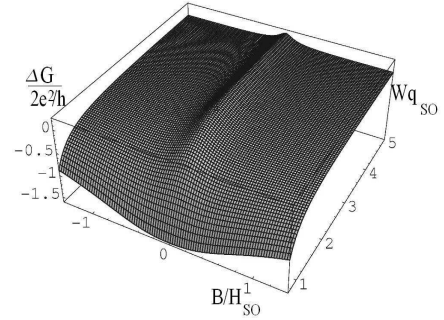,width= 6cm}
\vspace{-.3cm}
 \caption{
 The quantum conductivity correction  in units of $2 e^2/h$ as function of magnetic 
 field $B$ (scaled with  bulk relaxation field $H_{s}$), and the wire width $W$ (scaled with 
 spin-orbit length $L_{SO}$),  for pure Rashba coupling, $\delta_{SO}=1$.} \label{Fig5}
 \vspace{-1cm}
\end{center}
\end{figure}
 In the limit $W Q_{SO} \gg 1$ 
  the gaps of the triplet mode dispersions given in  Eq. (\ref{e0})
    coincide with the  2D gap values $\hbar D Q_{SO}^2 (1/2,1/2,1)$ of Eqs. (\ref{triplet2D})  (Note that the spin quantization axis is rotated by the unitary
    transformation).
   For  $W Q_{SO} < 1$ 
   the spin-orbit gap of the triplet mode $E_{T 0}$ is
    to first order in $t_{SO}$ and $c_{SO}$ given by  $\Delta_0 = D Q_{SO}^2 (2 c_{SO} \delta_{SO}^2 (1-\delta_{SO}^2) + t_{SO} \delta_{SO}^4/2)$
    and the gap of  $E_{T \pm}$ is $\Delta_{\pm} = \Delta_0/2 + D Q_{SO}^2 (2 c_{SO} - t_{SO}/2) \delta_{SO}^4$.
 Thus,   for  $W Q_{SO} \ll 1$   the weak localization correction 
    is  
    \begin{eqnarray} \label{wl1D}
 && \Delta \sigma =   \frac{\sqrt{H_W}}{\sqrt{H_{\varphi}+ B^*(W)/4}} - \frac{\sqrt{H_W}}{\sqrt{ H_{\varphi} 
 + B^*(W)/4 +  H_{s}(W) }} 
 \nonumber \\  &&  
 - 2 \frac{\sqrt{H_W}}{ \sqrt{H_{\varphi}+ B^*(W)/4 + H_{s}(W)/2}},
 \end{eqnarray}
  in units of $e^2/h$.
  We defined $H_W = \hbar/(4 e W^2)$, and  the effective external magnetic field,
  \begin{equation} \label{beff}
   B^*(W) =  (1-1/(1+\frac{W^2}{3 l_B^2})) B.
   \end{equation}   
    The spin relaxation  field $H_{s}(W)$ is  for $W < L_{SO}$,
\begin{equation} \label{hso0}
 H_s (W) = \frac{1}{12}(\frac{W}{L_{SO}})^2 \delta_{SO}^2 H_{s},
\end{equation}
    suppressed in proportion to 
      $(W/L_{SO})^2$. The   analogy to 
  the  effective magnetic field, 
 Eq. (\ref{beff}), could be expected, since 
  the spin orbit coupling enters  the Cooperon, Eq. (\ref{cooperon2}),
  like an effective  magnetic vector potential\cite{falko}.
   Cubic Dresselhaus coupling gives rise
        to an additional spin relaxation term, Eq. (\ref{hgamma}),  which has no  analogy to a magnetic field and is therefore not suppressed.  
   When $W$ is larger than   spin-orbit length $L_{SO}$, 
     coupling to higher transverse modes becomes
    relevant\cite{aleiner}.  
       One  can expect that in  ballistic wires,
 $l_e>W$,  the spin relaxation rate is suppressed  
 in analogy  to the flux cancellation effect, which yields the 
weaker   rate, $1/ \tau_s = (W/C l_e) (D W^2/ 12 L_{SO}^4)$,
 where $C= 10.8$\cite{km02}.
         
  In conclusion,  in wires whose width $W$ is smaller than  bulk 
  spin orbit  length 
  $L_{SO}$  spin relaxation  due to linear Rashba and Dresselhaus spin-orbit coupling is suppressed.
    The
      spin  relaxes then 
       due to  small cubic Dresselhaus coupling, only. 
   Thus, the total spin relaxation rate as function of wire width is  for $W < L_{SO}$, 
    \begin{equation}\label{totaltaus}
    \frac{1}{ \tau_s} (W) =  \frac{1}{12}(\frac{W}{L_{SO}})^2 \delta_{SO}^2 \frac{1}{ \tau_s} + D \frac{(m_e^2 \epsilon_F \gamma)^2}{\hbar^3},
    \end{equation}
    where $1/\tau_{s} = 2 p_F^2  (\alpha_2^2 + (\alpha_1 - m_e \gamma \epsilon_F/2)^2 ) \tau  $. 
    The enhancement of  spin relaxation 
       length $L_s = \sqrt{D \tau_s (W)}$ can be understood      as follows: 
      The area an electron 
       covers by diffusion in  time $\tau_s$ is  $W L_s$.
       This  should be equal to the corresponding 2D area $L_{SO}^2$ \cite{falko}, which  yields $1/L_s^2 \sim 
        (W/L_{SO})^2/L_{SO}^2$, in agreement with Eq. (\ref{hso0}). 
      At  lower temperatures, when   dephasing length $L_{\varphi}$ exceeds 
       $L_{\gamma} = \hbar/m_e^2 \epsilon_F \gamma$,  a weak antilocalization
        peak 
       is recovered at small magnetic fields,   $l_B > L_{\gamma}$. 
   Reduction of  spin relaxation  has recently been observed     in optical measurements of n-doped  InGaAs quantum wires\cite{holleitner}, where $\delta_{SO} \approx 1$,  and in transport measurements\cite{hu05,gh05}, and also  in GaAs wires\cite{lshh04}. Ref. \cite{holleitner}
    reports   saturation of  spin relaxation  in narrow wires, $W \ll L_{SO}$, attributed to  cubic Dresselhaus coupling, in full agreement with      Eq. (\ref{totaltaus}). 
    
    We thank V. L. Fal'ko, and F. E. Meijer for stimulating discussions,
     I. Aleiner, C. Marcus, T. Ohtsuki, K. Slevin, K. Dittmer,  J. Ohe, and A. Wirthmann for
    helpful  discussions, and A. Chudnovskiy, K. Patton and E. Mucciolo for 
     useful comments. 
     We gratefully acknowledge the hospitality of 
     MPIPKS in Dresden, the Physics Department of Sophia University, Tokyo and  Aspen Center for Physics. 
     This work was supported by SFB508 B9.


\begin{references}




\bibitem{review} B. L. Altshuler, A. G. Aronov, D. E. Khmelnitskii, and A. I. Larkin, in Quantum Theory of Solids, edited by I. M. Lifshits (Mir, Moscow, 1982); G. Bergmann, Phys. Rep. {\bf 107}, 1 (1984); 	S. Chakravarty, A. Schmid, Phys. Rep.  {\bf 140},  193 (1986).


\bibitem{perel} M. I. D'yakonov and V. I. Perel', Sov. Phys. Solid State {\bf 13}, 3023 (1972).


\bibitem{kiselev} A. A. Kiselev, and K. W. Kim, Phys. Rev. B {\bf 61}, 13115 (2000).

\bibitem{meyer} J. S. Meyer, V. I. Fal'ko, B.L. Altshuler,
Nato Science Series II,  {\bf 72}, 117, ed. I.V. Lerner, B.L. Altshuler, V.I. Fal'ko, and T. Giamarchi (Kluwer Academic Publishers, Dordrecht, 2002).

\bibitem{holleitner} 
  A. W. Holleitner, V. Sih, R. C. Myers,
  A. C. Gossard, D. D.   	Awschalom,
  Phys. Rev. Lett. {\bf 97}, 036805 (2006).

\bibitem{hu05} A. Wirthmann, Y. S. Gui, C. Zehnder, C. Heyn,  D. Heitmann,C. -M. Hu, and S. Kettemann, Physica E {\bf 34},  493 (2006).

\bibitem{gh05} F. E. Meijer, private communication (2005).

\bibitem{schaepers} Th. Sch\" apers, V. A. Guzenko, M. G. Pala, U. Z\" ulicke, M. Governale, J. Knobbe, and H. Hardtdegen
 Phys. Rev. B {\bf 74}, 081301(R) (2006).

\bibitem{lshh04} R. Dinter, S. L\" ohr, S. Schulz, Ch. Heyn, and W. Hansen, unpublished (2005).

\bibitem{dasdatta} B. Datta and S. Das, Appl. Phys. Lett. {\bf 56}, 665 (1990).

\bibitem{spintronics} I. Zutic', J. Fabian, and S. Das Sarma
Rev. Mod. Phys. {\bf 76}, 323 (2004).

\bibitem{nagaoka} S. Hikami, A. I. Larkin, and Y. Nagaoka,
  Prog. Theor. Phys. {\bf 63 }, 707 (1980). 

\bibitem{dresselhaus} G. Dresselhaus, Phys. Rev. {\bf 100}, 580(1955). 

\bibitem{rashba} E. I. Rashba, Fiz. Tverd. Tela (Leningrad) {\bf 2}, 1224 (1960)
[Sov. Phys. Solid State {\bf 2}, 1109 (1960)]; Yu. A. Bychkov
and E. I. Rashba, Pis'ma Zh. Eksp. Teor. Fiz. {\bf 39}, 66
(1984).

\bibitem{knap} W. Knap,
C. Skierbiszewski, A. Zduniak, E. Litwin-Staszewska,
D. Bertho, F. Kobbi, J. L. Robert, G. E. Pikus, F. G. Pikus, S. V.
Iordanskii, V. Mosser, K. Zekentes, and Yu. B. Lyanda-Geller,
 Phys. Rev. B {\bf 53}, 3912 (1996). 

\bibitem{miller} J. B. Miller,
D.M. Zumbuhl, C.M. Marcus,
Y.B. Lyanda-Geller, D. Goldhaber-Gordon, K. Campman, and A.C.
Gossard,
 Phys. Rev. Lett. {\bf 90}, 076807
  (2003). 


\bibitem{lg98} Y. Lyanda-Geller, Phys. Rev. Lett. {\bf 80}, 4273 (1998).  

\bibitem{golub}  L. E. Golub, Phys. Rev. B {\bf 71}, 235310  (2005). 

\bibitem{af01} I. L. Aleiner, V. I. Fal'ko, Phys. Rev. Lett. {\bf 87},  256801 (2001). 


\bibitem{iordanskii} S. V. Iordanskii, Yu. B. Lyanda-Geller, and G. E. Pikus,
JETP Lett. {\bf 60}, 206 (1994).


\bibitem{kurdak} C. Kurdak, A. M. Chang, A. Chin, and T. Y. Chang,
 Phys Rev. B {\bf 46}, 6846 (1992). 
 


\bibitem{falko} V. L. Fal'ko,  private communication, (2003). 

\bibitem{aleiner} I. L. Aleiner, privtate communication (2006). 

\bibitem{km02} V. K. Dugaev, D. E. Khmeln'itskii, JETP \textbf{59},
1038 (1984); C. W. J. Beenakker and H. van Houten, Phys. Rev. B
\textbf{37}, 6544 (1988); S. Kettemann and R. Mazzarello, Phys. Rev.
B \textbf{65}, 085318 (2002).


 
 
 
\end{references}
\end{document}